\newcommand{\CLASSINPUTtoptextmargin}{19mm}%
\newcommand{\CLASSINPUTbottomtextmargin}{43mm}%
\newcommand{\CLASSINPUTinnersidemargin}{12.9mm}%
\newcommand{\CLASSINPUToutersidemargin}{12.9mm}%
\documentclass[conference,10pt,a4paper]{IEEEtran}
\usepackage{amsmath}
\usepackage{cite}
\usepackage{times}
\usepackage{graphicx}
\usepackage{makecell}  
\usepackage{multirow}
\usepackage[none]{hyphenat}
\usepackage{float}
\usepackage{subfig}
\usepackage{iftex}
\usepackage{hyperref}  
\usepackage{xcolor}
\ifPDFTeX%
\usepackage{t1enc}
\usepackage{times}
\fi%
\ifXeTeX%
\usepackage{fontspec}
\fi%
\ifLuaTeX%
\usepackage{fontspec}
\fi%
\makeatletter

\def\@maketitle{\newpage
\bgroup\par\addvspace{0.5\baselineskip}\centering%
\ifCLASSOPTIONtechnote
   {\bfseries\large\@IEEEcompsoconly{\sffamily}\@title\par}\vskip 1.3em{\lineskip .5em\@IEEEcompsoconly{\sffamily}\@author
   \@IEEEspecialpapernotice\par{\@IEEEcompsoconly{\vskip 1.5em\relax
   \@IEEEtitleabstractindextextbox{\@IEEEtitleabstractindextext}\par
   \hfill\@IEEEcompsocdiamondline\hfill\hbox{}\par}}}\relax
\else
   \vskip0.2em{\EuMWtitlesize\ifCLASSOPTIONtransmag\bfseries\LARGE\fi\@IEEEcompsoconly{\sffamily}\@IEEEcompsocconfonly{\normalfont\normalsize\vskip 2\@IEEEnormalsizeunitybaselineskip
   \bfseries\Large}\@title\par}\vskip1.0em\par
   \ifCLASSOPTIONconference%
      {\@IEEEspecialpapernotice\mbox{}\vskip\@IEEEauthorblockconfadjspace%
       \mbox{}\hfill\begin{@IEEEauthorhalign}\@author\end{@IEEEauthorhalign}\hfill\mbox{}\par}\relax
   \else
      \ifCLASSOPTIONpeerreviewca
         {\@IEEEcompsoconly{\sffamily}\@IEEEspecialpapernotice\mbox{}\vskip\@IEEEauthorblockconfadjspace%
          \mbox{}\hfill\begin{@IEEEauthorhalign}\@author\end{@IEEEauthorhalign}\hfill\mbox{}\par
          {\@IEEEcompsoconly{\vskip 1.5em\relax
           \@IEEEtitleabstractindextextbox{\@IEEEtitleabstractindextext}\par\hfill
           \@IEEEcompsocdiamondline\hfill\hbox{}\par}}}\relax
      \else
         \ifCLASSOPTIONtransmag
           {\@IEEEspecialpapernotice\mbox{}\vskip\@IEEEauthorblockconfadjspace%
            \mbox{}\hfill\begin{@IEEEauthorhalign}\@author\end{@IEEEauthorhalign}\hfill\mbox{}\par
           {\vspace{0.5\baselineskip}\relax\@IEEEtitleabstractindextextbox{\@IEEEtitleabstractindextext}\vspace{-1\baselineskip}\par}}\relax
         \else
           {\lineskip.5em\@IEEEcompsoconly{\sffamily}\sublargesize\@author\@IEEEspecialpapernotice\par
           {\@IEEEcompsoconly{\vskip 1.5em\relax
            \@IEEEtitleabstractindextextbox{\@IEEEtitleabstractindextext}\par\hfill
            \@IEEEcompsocdiamondline\hfill\hbox{}\par}}}\relax
         \fi
      \fi
   \fi
\fi\par\addvspace{0.0\baselineskip}\egroup}

\def\EuMWtitlesize{\@setfontsize{\EuMWtitlesize}{24}{24pt}}
\def\EuMWauthorsize{\@setfontsize{\EuMWauthorsize}{11}{11pt}}
\def\EuMWaffilsize{\@setfontsize{\EuMWaffilsize}{10}{10pt}}
\def\EuMWcaptionsize{\@setfontsize{\EuMWcaptionsize}{9}{10pt}}
\def\EuMWbibsize{\@setfontsize{\EuMWbibsize}{8}{10pt}}

\def\@IEEEauthorblockNstyle{\EuMWauthorsize\@IEEEcompsocnotconfonly{\sffamily}\@IEEEcompsocconfonly{\large}}
\def\@IEEEauthorblockAstyle{\EuMWaffilsize\@IEEEcompsocnotconfonly{\sffamily}\@IEEEcompsocconfonly{\itshape}\@IEEEcompsocconfonly{\large}}
\def\@IEEEauthordefaulttextstyle{\EuMWauthorsize\@IEEEcompsocnotconfonly{\sffamily}\sublargesize}

\def\thebibliography#1{\section*{\refname}%
    \addcontentsline{toc}{section}{\refname}%
    \EuMWbibsize\@IEEEcompsocconfonly{\small}\vskip 0.3\baselineskip plus 0.1\baselineskip minus 0.1\baselineskip
    \list{\@biblabel{\@arabic\c@enumiv}}%
    {\settowidth\labelwidth{\@biblabel{#1}}%
    \leftmargin\labelwidth
    \advance\leftmargin\labelsep\relax
    \itemsep \IEEEbibitemsep\relax
    \usecounter{enumiv}%
    \let\p@enumiv\@empty
    \renewcommand\theenumiv{\@arabic\c@enumiv}}%
    \let\@IEEElatexbibitem\bibitem%
    \def\bibitem{\@IEEEbibitemprefix\@IEEElatexbibitem}%
\def\newblock{\hskip .11em plus .33em minus .07em}%
\ifCLASSOPTIONtechnote\sloppy\clubpenalty4000\widowpenalty4000\interlinepenalty100%
\else\sloppy\clubpenalty4000\widowpenalty4000\interlinepenalty500\fi%
    \sfcode`\.=1000\relax}

\long\def\@makecaption#1#2{%
\ifx\@captype\@IEEEtablestring%
\par\@IEEEtabletopskipstrut
\else
\@IEEEfigurecaptionsepspace
\fi
\setbox\@tempboxa\hbox{\normalfont\footnotesize {#1.}\nobreakspace\nobreakspace #2}%
\ifdim \wd\@tempboxa >\hsize%
\setbox\@tempboxa\hbox{\normalfont\footnotesize {#1.}\nobreakspace\nobreakspace}%
\parbox[t]{\hsize}{\normalfont\footnotesize\noindent\unhbox\@tempboxa#2}%
\else
\ifCLASSOPTIONconference \hbox to\hsize{\normalfont\footnotesize\hfil\box\@tempboxa\hfil}%
\else \hbox to\hsize{\normalfont\footnotesize\box\@tempboxa\hfil}%
\fi\fi
\ifx\@captype\@IEEEtablestring%
\@IEEEtablecaptionsepspace
\else
\fi}

\newlength\tablecaptiontotableskip
\newlength\figuretocaptionskip
\def\@IEEEfigurecaptionsepspace{\vskip\figuretocaptionskip\relax}%
\def\@IEEEtablecaptionsepspace{\vskip\tablecaptiontotableskip\relax}%

\def\abstract{\normalfont%
\@IEEEabskeysecsize\bfseries\textit{\abstractname}\,\bfseries\textit{---}\,%
\@IEEEgobbleleadPARNLSP}%

\def\IEEEkeywords{\normalfont%
\@IEEEabskeysecsize\bfseries\textit{\IEEEkeywordsname}\,\bfseries\textit{---}\,%
\@IEEEgobbleleadPARNLSP}%
\def\endIEEEkeywords{\relax\vspace{0.67ex}%
\par\if@twocolumn\else\endquotation\fi%
\normalsize\normalfont}%

\def\@IEEEauthorblockNtopspace{0ex}
\def\@IEEEauthorblockAtopspace{1mm}
\def\tablename{Table}
\def\thetable{\arabic{table}}
\def\IEEEkeywordsname{Keywords}
\def\subsubsection{\@startsection{subsubsection}{3}{\z@}{1.5ex plus 1.5ex minus 0.5ex}%
{0.7ex plus .5ex minus 0ex}{\normalfont\normalsize\itshape}}%
\newlength{\CPheadmatchindent}%
\def\@seccntformat#1{\hbox to\CPheadmatchindent{\csname the#1dis\endcsname}\hskip 0.1em \relax}
\IEEEilabelindentA \parindent
\IEEEilabelindent \IEEEilabelindentA
\IEEEelabelindent \parindent
\IEEEdlabelindent \parindent
\IEEElabelindent \parindent
\makeatother

\begin{document}
\raggedbottom
%
%
%
\title{BladeSEI: A Multi-Channel and Multi-Day Dataset for Robust Specific Emitter Identification of Radars}
%
%
\author{%
\IEEEauthorblockN{%
Ryan White\IEEEauthorrefmark{1}, 
Robert Jones\IEEEauthorrefmark{2}, 
Hassan Habib\IEEEauthorrefmark{2}, 
Matthew A. Ritchie\IEEEauthorrefmark{1}, 
Victoria Nockles\IEEEauthorrefmark{2}, 
Adam Hosford\IEEEauthorrefmark{3}
}

\IEEEauthorblockA{\IEEEauthorrefmark{1}Department of Electronic and Electrical Engineering,\\
University College London, WC1E 6BT, London, UK.\\
ryan.white.23@ucl.ac.uk}

\IEEEauthorblockA{\IEEEauthorrefmark{2}Defence Artificial Intelligence Research,\\
The Alan Turing Institute, NW1 2DB, London, UK.\\}

\IEEEauthorblockA{\IEEEauthorrefmark{3}Defence Science and Technology Laboratory (Dstl),\\
SP4 0JQ, Salisbury, UK.}
}%

\maketitle

\begin{abstract}

This paper introduces "BladeSEI", a multi-channel receiver, multi-day dataset of radar pulses recorded over-the-air from 12 Software Defined Radios (SDRs) for the purpose of Specific Emitter Identification (SEI). The aim of this dataset is to enable investigation into robust SEI techniques that are able to generalise across different receiver channels and recording sessions.

The dataset consists of two waveform types per emitter, a 50 microsecond chirped pulse of 40 MHz bandwidth and a 10 microsecond unmodulated pulse. Pulses are captured simultaneously on four synchronised analogue-to-digital converter (ADC) channels of a multi-channel receiver over three recording sessions on two days: a baseline session, an identical repeat on the following day, and a session in which the emitters were thermally conditioned to alter their operating temperature.

Initial results using Principal Component Analysis (PCA), Inception and ResNet based classifiers, trained and tested on a single ADC channel and recording session, show that the BladeRF emitters can be discriminated at high signal-to-noise ratio (SNR), with ResNet achieving a macro-F1 above 0.98 at 25 dB for both waveforms, and performance degrading towards random guessing at 0 dB.

\end{abstract}

\section{Introduction}

The electromagnetic (EM) spectrum is becoming increasingly congested due to the growing number of active emitters and their occupation of increasingly wider bandwidths. This significantly complicates the task of emitter detection and separation for electronic support measures (ESM) systems, especially when emitters share similar parameters.

Pulse de-interleaving is one technique that aims to separate an interleaved stream of received pulses into distinct pulse trains, each corresponding to an individual emitter. Conventional approaches typically exploit expert features such as carrier frequency, pulse width, pulse repetition interval (PRI), power and modulation characteristics to identify patterns that distinguish one emitter from another.

The increasing adoption of software-defined radar systems presents a significant challenge to conventional de-interleaving techniques. Such systems can change parameters on a pulse-to-pulse basis, causing expert-feature based de-interleaving algorithms to fail. Furthermore, a single software-defined platform could potentially emulate multiple different radars operating concurrently. Such capabilities can complicate ESM processing by increasing the apparent number of emitters and further adding to the difficulty of tracking emitters.

Specific emitter identification (SEI), also referred to as RF fingerprinting, addresses the problem of identifying emitters by exploiting distinctive radio-frequency (RF) fingerprints arising from unintentional variations in transmitter hardware and their effects on the transmitted waveform. These characteristics may be difficult to replicate or spoof, allowing SEI to support pulse de-interleaving where conventional approaches struggle and potentially detect emissions inconsistent with the expected characteristics of a known emitter. However, these benefits depend on the reliable recognition of emitter-specific characteristics across varying waveforms and operating conditions.

A fundamental challenge is that an observed RF fingerprint is not determined solely by the transmitting hardware. The received signal is also influenced by the environment (e.g. multipath, atmosphere) and receiver hardware (e.g. RF front-end, antenna). Consequently, a model may inadvertently learn receiver-specific or session-specific artefacts rather than characteristics intrinsic to the emitter. This can limit its ability to generalise to signals acquired using different receiving hardware or at different times (Also referred to as the "day-after-tomorrow problem"). Therefore, establishing robust SEI capabilities requires a dataset that enables evaluation across these varying constraints.

Research into RF fingerprinting has benefited substantially from publicly available datasets of commercial off-the-shelf (COTS) communications devices \cite{wisig}\cite{oracle}\cite{lorarffi}\cite{uavsig}, owing in part to the accessibility and relative affordability of these systems. The lack of experimentally acquired, publicly available radar datasets consequently limits the reproducibility of research and the ability to conduct meaningful comparisons between competing SEI techniques for radar waveforms. To address this gap, we introduce "BladeSEI", a publicly available dataset of radar pulses captured experimentally over the air using a Radio Frequency System-on-Chip (RFSoC) based receiver. The dataset incorporates recordings collected across multiple ADC channels of the RFSoC and multiple recording sessions, enabling investigation of the effects of receiver channel and temporal variation on radar SEI performance.

BladeSEI is intended to provide a common reference for evaluating and comparing radar SEI techniques, while supporting research into methods that generalise across receiver channels and recording sessions. By enabling systematic investigation of these sources of variability, the dataset aims to help determine whether SEI models learn robust emitter-specific features or rely on incidental characteristics of the acquisition process. In doing so, BladeSEI seeks to bridge the gap between controlled experimental evaluations and the demands of operational ESM systems.

The remainder of the paper is structured as follows, Section~\ref{sec:dataset} describes the dataset structure, hardware used, capture serials and the provided code, Section~\ref{sec:approach} describes some baseline results, Section~\ref{sec:code} describes the dataset licensing and provides download links to the code and dataset, and finally Section~\ref{sec:Conclusions} concludes the work.

\section{Dataset Description}\label{sec:dataset}

\subsection{Transmitter Description}\label{sec:hardware_tx}

The 12 transmitters used in the generation of this dataset were Nuand BladeRF 2.0 micro xA5 \cite{nuand} software-defined radios. The BladeRF is built around an Intel Cyclone V Field Programmable Gate Array (FPGA) connected to an Analog Devices AD9361 RF transceiver \cite{ad9361}, which provides the RF front-end and conversion between complex baseband in-phase (I) and quadrature (Q) samples and RF signals (and vice-versa).

Only a single transmit channel was used on each device during data collection. The AD9361 digital-to-analogue converter (DAC) was configured with a sample rate of 60 MS/s and transmitted at a centre frequency of 2.45 GHz, placing signals at the centre of the 2.4 GHz ISM (Industrial, Scientific and Medical) band. An RF power amplifier and dish antenna were used to transmit the BladeRFs signal over the air, the exact components are listed below in Table \ref{tab:DAC_channels}.

\begin{table}[h!]
    \centering
    \caption{Blade Transmitter Configurations}
    \label{tab:DAC_channels}
    \begin{tabular}{|c|c|c|c|c|}
        \hline
        \textbf{Port} & \textbf{Antenna} & \textbf{Amplifier} & \textbf{Filter} & \textbf{Attenuation} \\ \hline
        TX0 & \makecell{L-COM\\HG2418DPD}  & \makecell{Mini-Circuits\\ZJL-4HG+}  & \makecell{N/A}  & \makecell{N/A} \\ \hline
    \end{tabular}
\end{table} 

\begin{figure}[htbp]
\begin{center}
\includegraphics[scale=0.13]{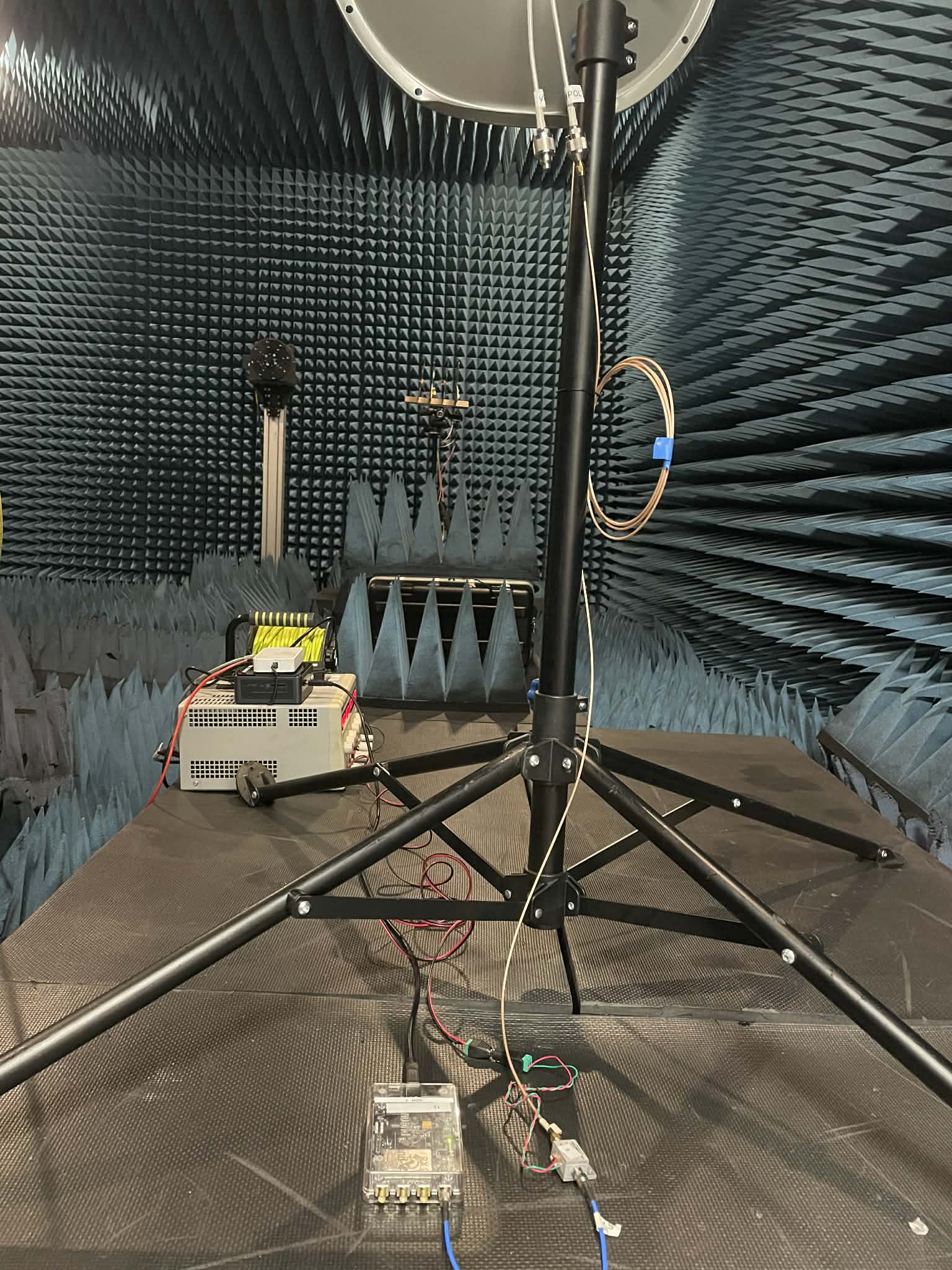}
\caption{Transmitter setup in the anechoic chamber, showing a BladeRF, transmit antenna, power supply and command and control computer}
\label{fig1}
\end{center}
\end{figure}

The BladeRFs were controlled by a host computer running \textit{GNU Radio}, which was responsible for setting the parameters (DAC sample rate and frequency) and streaming the pre-computed IQ samples plus associated padding from a circular buffer for transmission. This setup is shown in Fig. \ref{fig1}.

The padding was computed such that the pulses were transmitted at a pulse repetition frequency (PRF) of 1 kHz, mirroring the typical transmit duty-cycle of a real radar.

Two waveforms per emitter were used in the creation of the dataset to assess how pulse duration and bandwidth affect an emitter's fingerprint. The first is an unmodulated pulse of 10 $\mu$s duration transmitted at 2.46 GHz. The second is a chirped pulse of 50 $\mu$s duration and 40 MHz bandwidth, sweeping from 2.43 to 2.47 GHz.

Examples of both of these waveforms in both the time and frequency domain as seen by each of the receiver's ADCs is shown in Figs. \ref{fig6} and \ref{fig7} (Appendix). 

To increase the variability between the BladeRF emitters, the analogue bandwidth of each device was set to 20 MHz, pushing the start and end portions of the chirped signal outside the passband of the device's analogue filters. Because these filters are physical components, their responses differ between units, so operating at the edge of the passband amplifies the hardware-dependent fingerprint that distinguishes the emitters.

The hardware and software versions used for the BladeRF emitters during data collection were recorded to ensure reproducibility and are given below:

\begin{itemize}
  \item FX3 firmware: 2.6.0
  \item FPGA image version: 0.16.0
  \item GNU Radio version: 3.10.12
\end{itemize}

All BladeRF devices were purchased together as a single batch, providing the smallest variation in manufacturing induced hardware mismatches between devices, thus making the task of RF fingerprinting more difficult.

\subsection{Receiver Description}\label{sec:hardware_rx}

\begin{figure}[htbp]
\begin{center}
\includegraphics[scale=0.1]{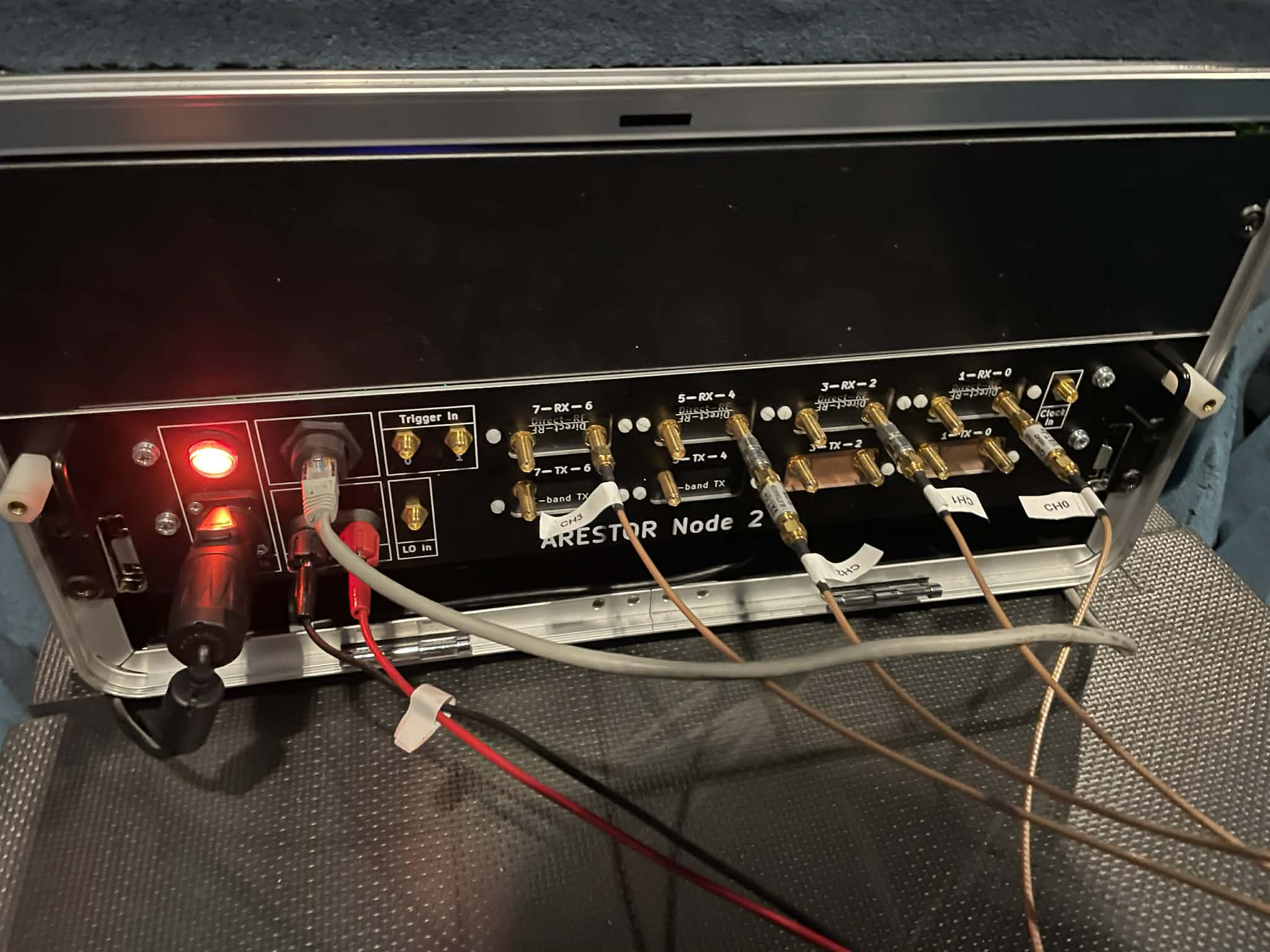}
\caption{Photograph of the front of the ARESTOR node, showing the power, network, and RF connections}
\label{fig2}
\end{center}
\end{figure}

The dataset was generated using signals received over-the-air in the 2.4 GHz ISM band using the UCL "ARESTOR" system shown in Fig. ~\ref{fig2}, which is an SDR based on an RFSoC architecture. ARESTOR has enabled experiments in active and passive radar \cite{9905049}, Integrated Sensing and Communications (ISAC) \cite{isac1}\cite{isac2}\cite{isac3} as well as Electronic Surveillance (ES) \cite{es1}\cite{es2}\cite{es3}. More details about the ARESTOR system and its hardware are provided in previous papers authored by the UCL Radar Group \cite{9784466} \cite{10149783}.

To generate the BladeSEI dataset, a single ARESTOR node acting as the receiver(s) was used to capture the pulses transmitted over-the-air by the BladeRFs. Four ADC channels, each implemented on a separate RF-Tile were used for receive in a "multi-tile synchronisation" (MTS) configuration.

The use of MTS enables independent RF tiles in the RFSoC, which may exhibit non-deterministic sample alignment, to have their relative data-path latency measured and corrected, enabling deterministic sample alignment across the ADC channels.

The receiver channels on the RFSoC were all set to operate at a sample rate of 3.84 GS/s, which is then decimated by a factor of 32 in two stages ($8\times$ by the RF Tile and $4\times$ by further FPGA processing) to achieve a final sample rate of 120 MS/s. As the desired signal is sampled in the 2nd Nyquist zone, spectral inversion occurs. The actual intermediate frequency after sampling is given in \eqref{eq:if_second_nyquist}. The digitised signals were then digitally down converted using a numerically controlled oscillator (NCO) operating at -1.39 GHz, which mixes the centre-frequency of our target signals to baseband.  

\begin{equation}
f_\text{IF} = \lvert f_s - f_\text{RF} \rvert, \quad \frac{f_s}{2} < f_\text{RF} < f_s
\label{eq:if_second_nyquist}
\end{equation}

\begin{figure}[htbp]
\begin{center}
\includegraphics[scale=0.133]{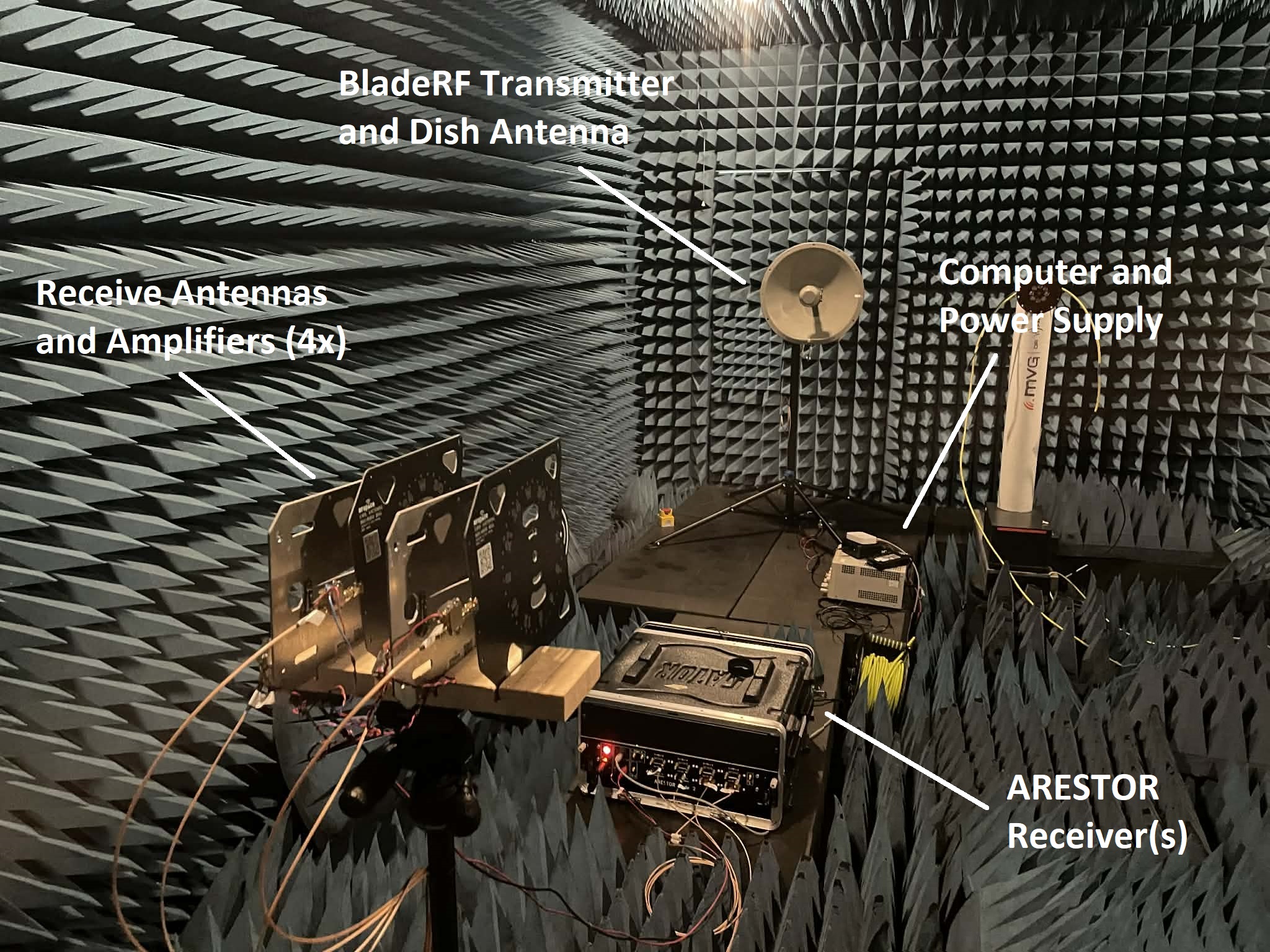}
\caption{Receiver setup in the anechoic chamber, showing the 4x co-located receive antennas and the ARESTOR node acting as a multi-channel receiver.}
\label{fig3}
\end{center}
\end{figure}

To enforce a greater degree of variability between the receive channels, making cross-ADC generalisation more difficult, different combinations of antennas (as seen in Fig.~\ref{fig3}) and amplifiers were used per channel as described in Table \ref{tab:ADC_channels}. Attenuators were used to ensure each channel has roughly the same SNR regardless of the gain and noise factor of the amplifier chosen.

\begin{table}[h!]
    \centering
    \caption{ADC Channel Configurations}
    \label{tab:ADC_channels}
    \begin{tabular}{|c|c|c|c|c|}
        \hline
        \textbf{Port} & \textbf{Antenna} & \textbf{Amplifier} & \textbf{Attenuation} \\ \hline
        ADC0 & \makecell{RFSPACE\\UWB3}    & \makecell{Mini-Circuits\\ZX60-123LN-S+}  
        &  4 dB  \\ \hline
        ADC2 & \makecell{RFSPACE\\TSA800}  & \makecell{Mini-Circuits\\ZX60-83LN12+ 1$^{\dagger}$} 
        & N/A  \\ \hline
        ADC4 & \makecell{RFSPACE\\UWB3}    & \makecell{Mini-Circuits\\ZX60-83LN12+ 2$^{\dagger}$}  
        &  1 dB \\ \hline
        ADC6 & \makecell{RFSPACE\\TSA800}  & \makecell{Mini-Circuits\\ZX60-3018G-S+}   
        & N/A \\ \hline
    \end{tabular}
    
\vspace{2pt}
\footnotesize{$^{\dagger}$ADC2 and ADC4 use two separate units of the same amplifier model.}

\end{table}

The receiver channels were configured to record for 4 seconds per acquisition, resulting in roughly 4000 digitised pulses per ADC, per acquisition.

\subsection{Experimental Captures}\label{sec:experiments}

Three sets of captures for the dataset were performed over two days, with all experiments conducted within an anechoic chamber. The first set was captured on the first day, with 24 acquisitions, covering the 12 BladeRF emitters with their two waveform modes each (short unmodulated pulse and longer frequency modulated pulse). The configuration of the transmitter and receivers in the chamber is shown in Fig. \ref{fig3}.

The second set was configured as an exact repeat of the first, but performed on the following day with no changes to the acquisition process or transmitter/receiver antenna geometries. This provides a controlled comparison for evaluating temporal variation in the observed transmitter fingerprints, with the experimental configuration held constant between the two sets.

The final set was also performed on the second day using the same geometries. Prior to acquisition, however, the BladeRF transmitters were placed on a 70 °C hot plate and allowed to transmit continuously for approximately five minutes before the receiver began digitising the pulses. This thermal conditioning was intended to alter the operating temperature of the transmitter hardware and consequently its RF characteristics, as temperature changes in components such as oscillators, analogue filters, amplifiers, and other elements of the RF signal chain may alter the transmitted waveform. This final set therefore provides a means of evaluating the sensitivity of the observed transmitter fingerprints to changes in the thermal operating condition of the emitters.

A thermal image, showing an unheated BladeRF emitter, a heated BladeRF during transmission and an idle BladeRF are shown below in Fig. \ref{fig4}. 

\begin{figure}[htbp]
\begin{center}
\includegraphics[scale=0.15]{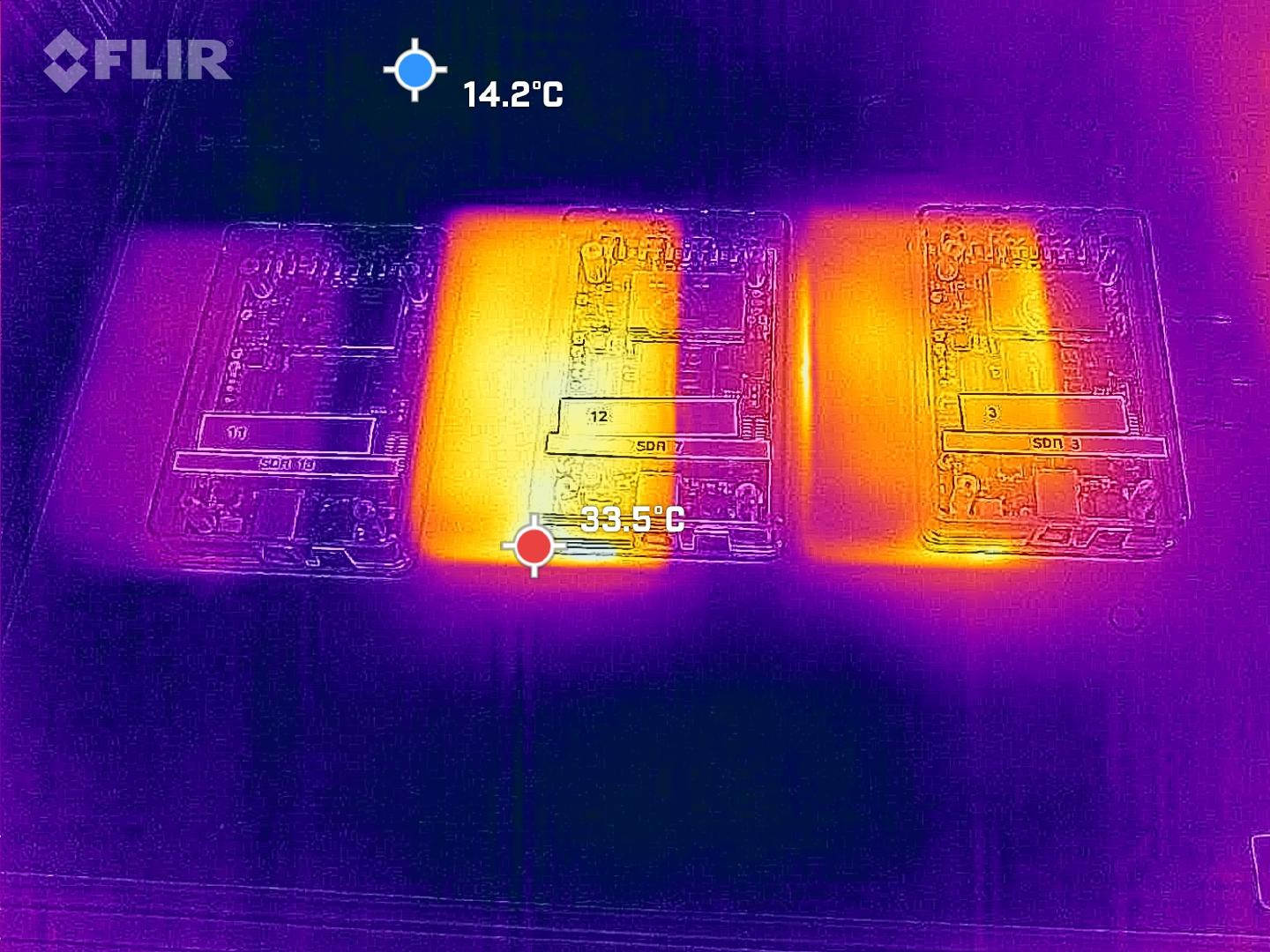}
\caption{Thermal image, showing a BladeRF without heating, a heated BladeRF after transmission and an idle BladeRF. The thermal overlay is slightly offset in the image.}
\label{fig4}
\end{center}
\end{figure}

\subsection{Post-Processing and Dataloader}\label{sec:dataset_code}

A basic CFAR (constant false alarm rate) based pulse detector is used to detect individual pulses from the continuous IQ data produced by each ADC and write them to \textit{HDF5} files in chunks of up to 20 pulses. Any pulse that is cut short by virtue of being at the start or end of a capture is discarded. Regardless of the waveform used, each pulse is stored as 8192 complex baseband samples by extending the window equally before and after the detected pulse using the surrounding noise samples to meet this target length. Each pulse is stored alongside a set of labels (capture day, environment, temperature, emitter and waveform type) and the capture metadata for each ADC.

The provided \textit{PyTorch} dataset class indexes a directory of \textit{HDF5} files containing the raw data and provides the ability to filter pulses by day, environment, temperature, emitter, waveform type and ADC. When more than one ADC is requested of the dataset class, the pulses can either be concatenated (increasing the dataset by a factor of the number of ADCs used) or stacked as multiple-views, enabling techniques such as contrastive loss to be used.

Stacking is enabled by comparing the measured time-of-arrival (ToA) of each pulse relative to the start of the capture. If a pulse arrives with a ToA within 2 $\mu$s between ADCs, these are considered to be the same physical pulse as digitised by each ADC independently.

Any pulse with a measured SNR below 31 dB is dropped, if the dataloader is set to provide stacked pulses across ADCs, the whole group is dropped.

Augmentations and SNR evaluation are handled by an "augmented" dataset wrapper, which provides additional preprocessing to the raw pulses provided by the base dataloader. Given a list of target SNRs, pulses are duplicated for the number of SNRs requested and noise is added to bring each duplicate to the desired SNR, allowing training and validation across a representative range of SNRs.

Additionally, random frequency and phase shifts can be applied to augment the data, or carrier-frequency offset (CFO) and phase variations can be corrected deterministically to reduce variability between captures. For deterministic correction, the CFO of each pulse is estimated, and its phase is aligned to a fixed reference anchor, defined as the first pulse in the dataset before any down-selection is performed.

Finally, the augmented wrapper provides functionality for the addition of a magnitude channel (given by $\sqrt{I^2 + Q^2}$), normalisation to peak magnitude and flattening of the tensor into a single vector (for PCA).

We provide training and evaluation \textit{Jupyter notebooks} for 1-dimensional ResNet and Inception style convolutional networks and for a PCA and $k$-nearest neighbour baseline. All use a repeatable seeded 80/20 train/test split. Trained model weights and emitter-to-index mappings are stored as checkpoints. For evaluation, the saved models are reloaded and tested on the test split at each SNR from 0 to 30 dB in 5 dB steps, producing a normalised confusion matrix and macro-F1 score at each SNR.

\section{Classification Performance}\label{sec:approach}

To test that it is possible to discriminate between the 12 BladeRF emitters using both waveforms (chirp and pulse), the example \textit{Jupyter notebooks} use both conventional and contemporary machine learning approaches as described in Section~\ref{sec:dataset_code} on ADC0 data only for the first recording session:

\begin{itemize}
    \item \textbf{PCA with a $k$-nearest neighbour classifier.} Each pulse is flattened into a single vector ($[I, Q, |z|]$, where $|z| = \sqrt{I^2 + Q^2}$), normalised by its peak magnitude), standardised, projected onto 50 principal components, and classified using $k{=}5$ neighbours.
    \item \textbf{Inception-based and ResNet-based 1-D CNNs.} Both take the same three-channel input as PCA ($[I, Q, |z|]$) and are trained with cross-entropy loss using AdamW (learning rate $10^{-4}$, weight decay $0.01$, cosine annealing to $10^{-5}$ over 50 epochs, batch size 64, early stopping with a patience of 15 epochs on the test loss, keeping the weights with the lowest test loss).
\end{itemize}

\begin{figure}[htbp]
\begin{center}
\includegraphics[scale=0.9]{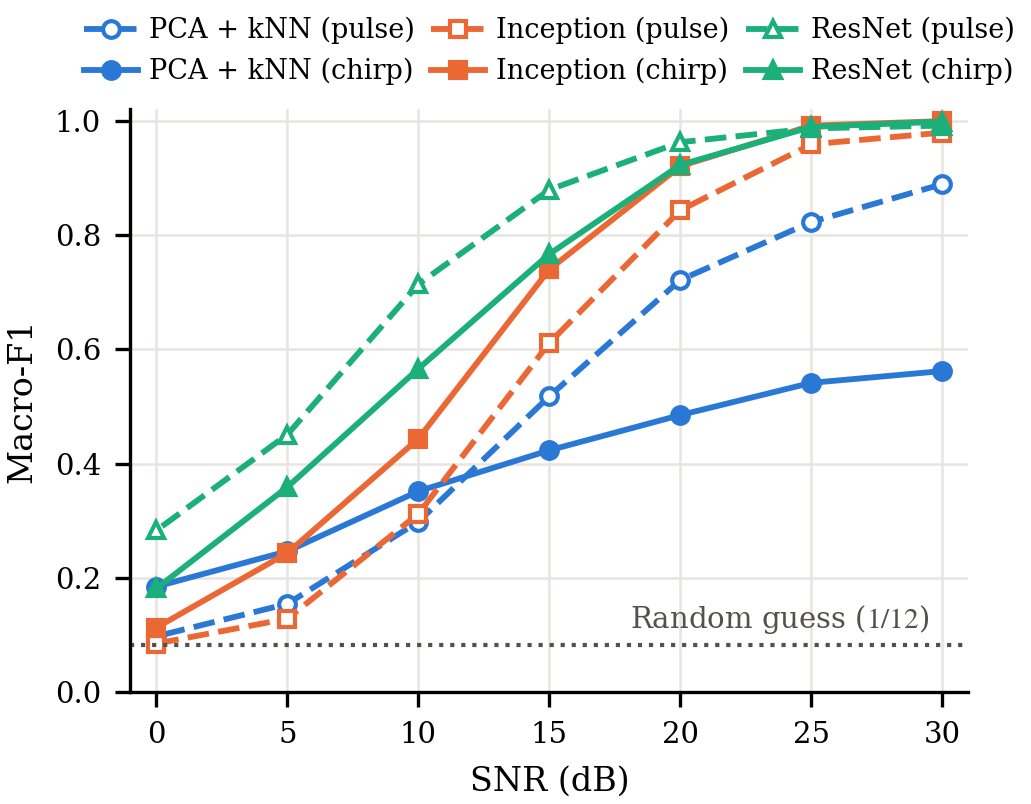}
\caption{Macro-F1 Score for each model/waveform combination against SNR.}
\label{fig5}
\end{center}
\end{figure}

Up to 1,000 pulses per emitter are used, giving up to 12,000 pulses per waveform, which are split 80/20 into training and test sets. Each training pulse is duplicated at seven target SNRs from 0 to 30 dB in 5 dB steps by adding white Gaussian noise, therefore each model sees 67,200 pulses in training across the specified SNR range.

For testing, each held-out pulse is evaluated separately at each SNR from 0 to 30 dB, and performance is reported as macro-F1 and normalised confusion matrices.

Fig.~\ref{fig5} shows macro-F1 against SNR for each classifier and waveform. Both convolutional networks exceed a macro-F1 of 0.95 at 25 dB and above for both waveforms. ResNet performs best as SNR decreases, achieving 0.71 (pulse) and 0.57 (chirp) at 10 dB, compared with 0.31 and 0.44 for Inception. PCA + kNN performs considerably worse, reaching only 0.56 on the chirp at 30 dB. Neither waveform is consistently easier to classify: Inception performs better on the chirp at every SNR, whereas ResNet performs better on the unmodulated pulse at 20 dB and below. At 0 dB all classifiers approach the random-guess level of $1/12$. Confusion matrices for each model/waveform combination at 10 dB SNR are shown in Fig.~\ref{fig8}

\section{Data and Code Availability}\label{sec:code}

The dataset and some source code to get started are publicly available, the links for which are provided below:

\begin{itemize}
    \item \textbf{Dataset:} The full dataset as described in this paper is available at 
    \url{https://doi.org/10.5522/04/34755795}. 

    \item \textbf{Example Processing Scripts:} The data-loader, visualisation and baseline machine learning scripts as described in Section \ref{sec:dataset_code} are available at \url{https://github.com/UCLRadarGroup/BladeSEI}. 

\end{itemize}

The dataset is released under the CC BY 4.0 license and the code is released under the GPL-3.0 license.

\section{Conclusions}\label{sec:Conclusions}


This paper has introduced "BladeSEI", an experimentally captured multi-channel receiver, multi-day dataset of two radar pulse modulations recorded over-the-air from 12 different emitters.

This dataset has been created to support the development and evaluation of specific emitter identification / RF fingerprinting techniques for radar waveforms that are able to generalise across receivers and recording sessions.

Baseline results on a single receiver channel and recording session show that the 12 emitters can be discriminated at high SNR for both waveforms, with a ResNet style network performing best.

Generalisation across receiver channels, recording days and thermal conditions is not evaluated in this paper and is left to future work.

\section*{Acknowledgments}\label{sec:acknowledgments}

This dataset was generated under an iCASE PhD Studentship funded by the Engineering and Physical Sciences Research Council (EPSRC) and the Defence Science and Technology Laboratory (Dstl). Prof. Ritchie’s time is supported by The Royal Academy of Engineering and Leonardo UK under the Research Chairs and Senior Research Fellowships programme.

Dr Robert Jones's and Dr Victoria Nockles's time was supported by the Turing’s Defence and Security programme through a partnership with the UK government in accordance with the framework agreement between His Majesty's Government (HMG) and The Alan Turing Institute.

Hassan Habib's time was supported by the Alan Turing Institute within the Turing Summer Internship programme 2026.


\bibliography{EuRad24_rhwhite}

@INPROCEEDINGS{9784466,
  author={Peters, Nial and Horne, Colin and Ritchie, Matthew A.},
  booktitle={2021 18th European Radar Conference (EuRAD)}, 
  title={{ARESTOR}: A Multi-role {RF} Sensor based on the {Xilinx RFSoC}}, 
  year={2022},
  volume={},
  number={},
  pages={102-105},
  doi={10.23919/EuRAD50154.2022.9784466}
}

@INPROCEEDINGS{10149783,
  author={Peters, Nial J. and Horne, Colin P. and Amiri, Amin D. and Beasley, Piers and Ritchie, Matthew A.},
  booktitle={2023 IEEE Radar Conference (RadarConf23)}, 
  title={Modular Multi-Channel RFSoC System Expansion and Array Design}, 
  year={2023},
  volume={},
  number={},
  pages={1-4},
  doi={10.1109/RadarConf2351548.2023.10149783}
}

@INPROCEEDINGS{9905049,
  author={Ritchie, M. and Peters, N. and Horne, C.},
  booktitle={2022 23rd International Radar Symposium (IRS)}, 
  title={Joint Active Passive Sensing using a Radio Frequency System-on-a-Chip based sensor}, 
  year={2022},
  volume={},
  number={},
  pages={130-135},
  doi={10.23919/IRS54158.2022.9905049}
}

@ARTICLE{isac1,
  author={Temiz, Murat and Horne, Colin and Peters, Nial J. and Ritchie, Matthew A. and Masouros, Christos},
  journal={IEEE Transactions on Microwave Theory and Techniques}, 
  title={An Experimental Study of Radar-Centric Transmission for Integrated Sensing and Communications}, 
  year={2023},
  volume={71},
  number={7},
  pages={3203-3216},
  doi={10.1109/TMTT.2023.3234309}
}

@INPROCEEDINGS{isac2,
  author={Temiz, Murat and Peters, Nial J. and Horne, Colin and Ritchie, Matthew A. and Masouros, Christos},
  booktitle={2023 IEEE Radar Conference (RadarConf23)}, 
  title={Radar-Centric ISAC Through Index Modulation: Over-the-air Experimentation and Trade-offs}, 
  year={2023},
  volume={},
  number={},
  pages={1-6},
  doi={10.1109/RadarConf2351548.2023.10149620}}

@INPROCEEDINGS{isac3,
  author={Winter, Robert S. C. and Catherall, Aled and Masouros, Christos and Ritchie, Matthew},
  booktitle={2024 IEEE International Workshop on Technologies for Defense and Security (TechDefense)}, 
  title={Impact and Mitigation of Index Modulation on Radar Quality for a FMCW-Based Communication System}, 
  year={2024},
  volume={},
  number={},
  pages={485-489},
  doi={10.1109/TechDefense63521.2024.10863567}
}

@ARTICLE{es1,
  author={Horne, Colin and Peters, Nial J. and Ritchie, Matthew A.},
  journal={IEEE Access}, 
  title={Classification of LoRa Signals With Real-Time Validation Using the Xilinx Radio Frequency System-on-Chip}, 
  year={2023},
  volume={11},
  number={},
  pages={26211-26223},
  doi={10.1109/ACCESS.2023.3252170}
}

@INPROCEEDINGS{es2,
  author={White, Ryan and Horne, Colin and Ritchie, Matthew A.},
  booktitle={2024 21st European Radar Conference (EuRAD)}, 
  title={An Empirical Evaluation of PDW Extraction Techniques Against Weak LPI Pulsed Radar}, 
  year={2024},
  volume={},
  number={},
  pages={260-263},
  doi={10.23919/EuRAD61604.2024.10734975}
}

@INPROCEEDINGS{es3,
  author={White, Ryan and Winter, Robert S. C. and Horne, Colin and Ritchie, Matthew A.},
  booktitle={2025 IEEE International Radar Conference (RADAR)}, 
  title={Evaluation of YOLO for Automatic Radar Waveform Detection and Classification in Congested EME}, 
  year={2025},
  volume={},
  number={},
  pages={1-6},
  doi={10.1109/RADAR52380.2025.11031686}
}

@online{nuand,
  author  = {{Nuand LLC}},
  title   = {{bladeRF 2.0 micro xA5}},
  url     = {https://www.nuand.com/product/bladerf-xa5/},
  urldate = {2026-09-02},
  year    = {2026}
}

@online{ad9361,
  author  = {{Analog Devices, Inc.}},
  title   = {{AD9361: 2 x 2 RF Agile Transceiver}},
  url     = {https://www.analog.com/media/en/technical-documentation/data-sheets/ad9361.pdf},
  urldate = {2026-09-02}
}

@article{wisig,
  title={WiSig: A Large-Scale WiFi Signal Dataset for Receiver and Channel Agnostic RF Fingerprinting}, 
  volume={10}, 
  ISSN={2169-3536},
  DOI={10.1109/ACCESS.2022.3154790}, 
  journal={IEEE Access}, 
  author={Hanna, Samer and Karunaratne, Samurdhi and Cabric, Danijela},
  year={2022}, 
  pages={22808–22818} 
}

@inproceedings{oracle,  
   title={ORACLE: Optimized Radio clAssification through Convolutional neuraL nEtworks},  
   author={Sankhe, Kunal and Belgiovine, Mauro and Zhou, Fan and Riyaz, Shamnaz and Ioannidis, Stratis and Chowdhury, Kaushik},  
   booktitle={IEEE INFOCOM 2019-IEEE Conference on Computer Communications},  pages={370--378},  
   year={2019},  
   organization={IEEE}

}

@data{uavsig,
  author = {Zhao, Tianyi and Domae, Benjamin and Steigerwald, Connor and Paradis, Luke B. and Chabuk, Tim and Cabric, Danijela},
  publisher = {UCLA Dataverse},
  title = {{UAVSig: Drone RF Signal Detection and Fingerprinting Dataset}},
  year = {2024},
  version = {V4},
  doi = {10.25346/S6/LVRRAE},
  url = {https://doi.org/10.25346/S6/LVRRAE}
}

@data{lorarffi,
  doi = {10.21227/qqt4-kz19},
  url = {https://dx.doi.org/10.21227/qqt4-kz19},
  author = {Guanxiong Shen and Junqing Zhang and Alan Marshall},
  publisher = {IEEE Dataport},
  title = {LoRa RFFI dataset},
  year = {2022}
}
\bibliographystyle{ieeetr}




\onecolumn
\section{Appendices}\label{sec:Appendices}

\begin{figure}[H]
\centering
\includegraphics[height=0.42\textheight,width=\textwidth,keepaspectratio]{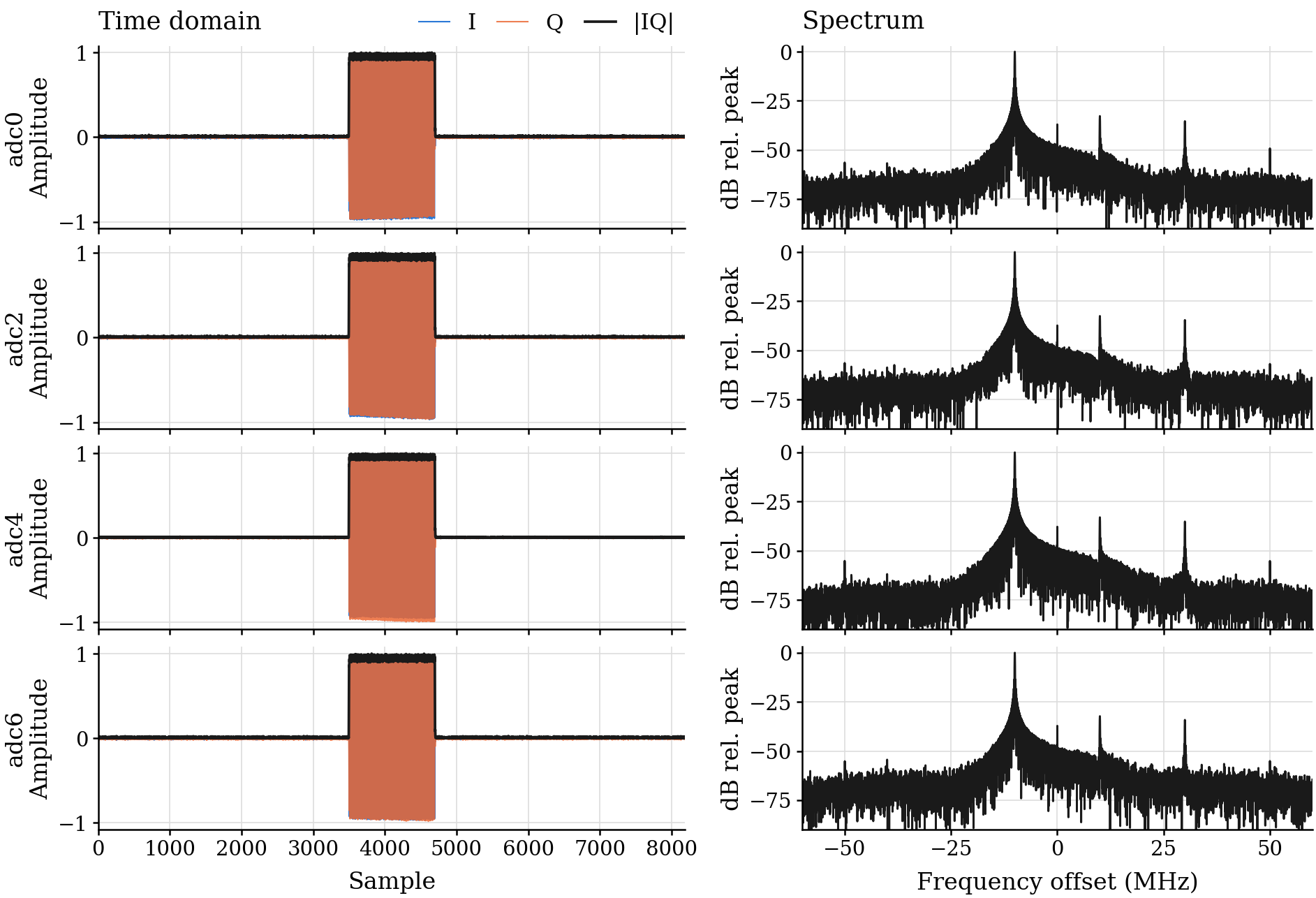}
\caption{Normalised time and frequency domain plots for the same unmodulated pulse digitised by the four ADCs}
\label{fig6}
\end{figure}

\vspace{2.5em}

\begin{figure}[H]
\centering
\includegraphics[height=0.42\textheight,width=\textwidth,keepaspectratio]{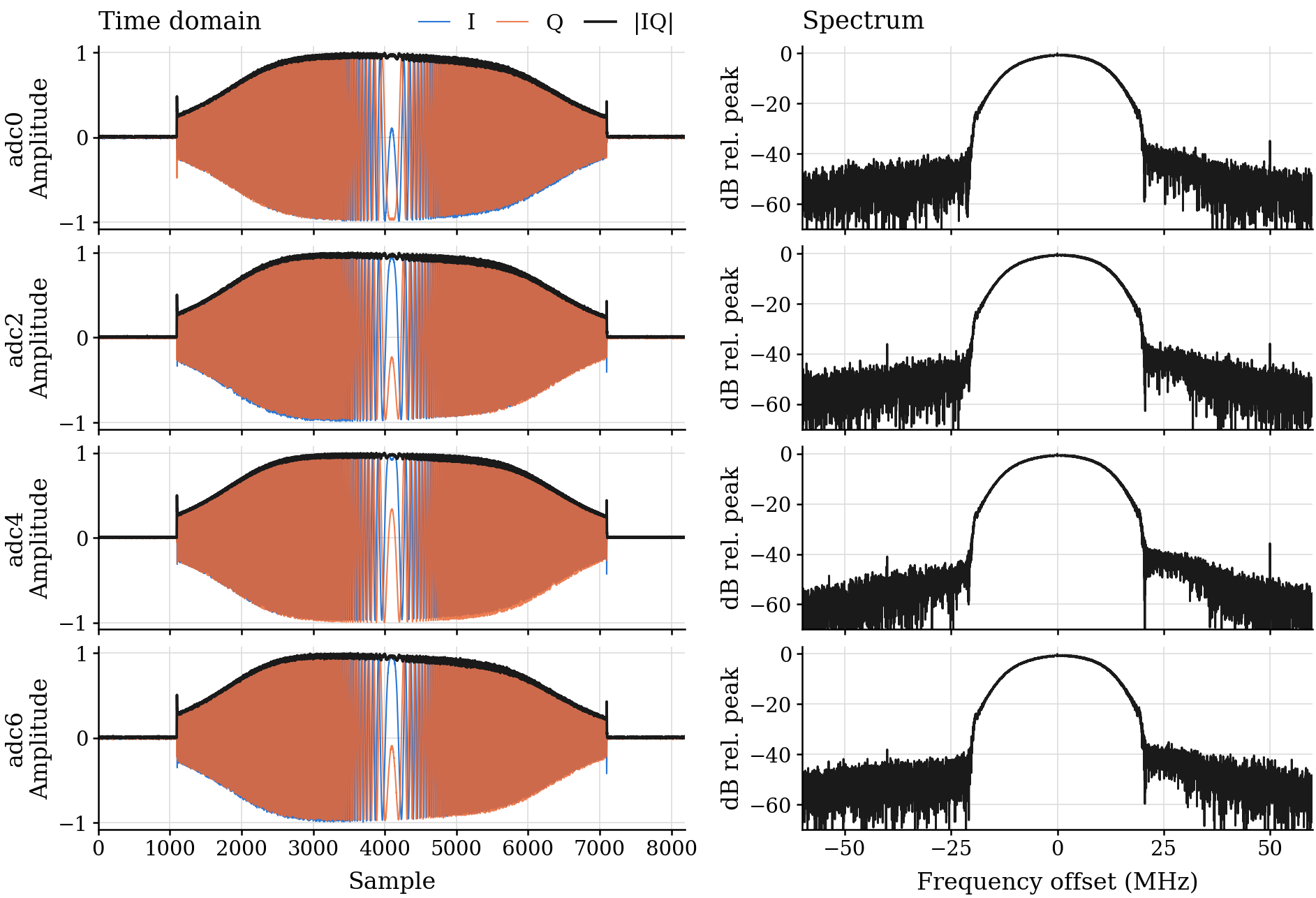}
\caption{Normalised time and frequency domain plots for the same chirped pulse digitised by the four ADCs}
\label{fig7}
\end{figure}


\begin{figure}[H]
\centering
\includegraphics[height=1\textheight,width=\textwidth,keepaspectratio]{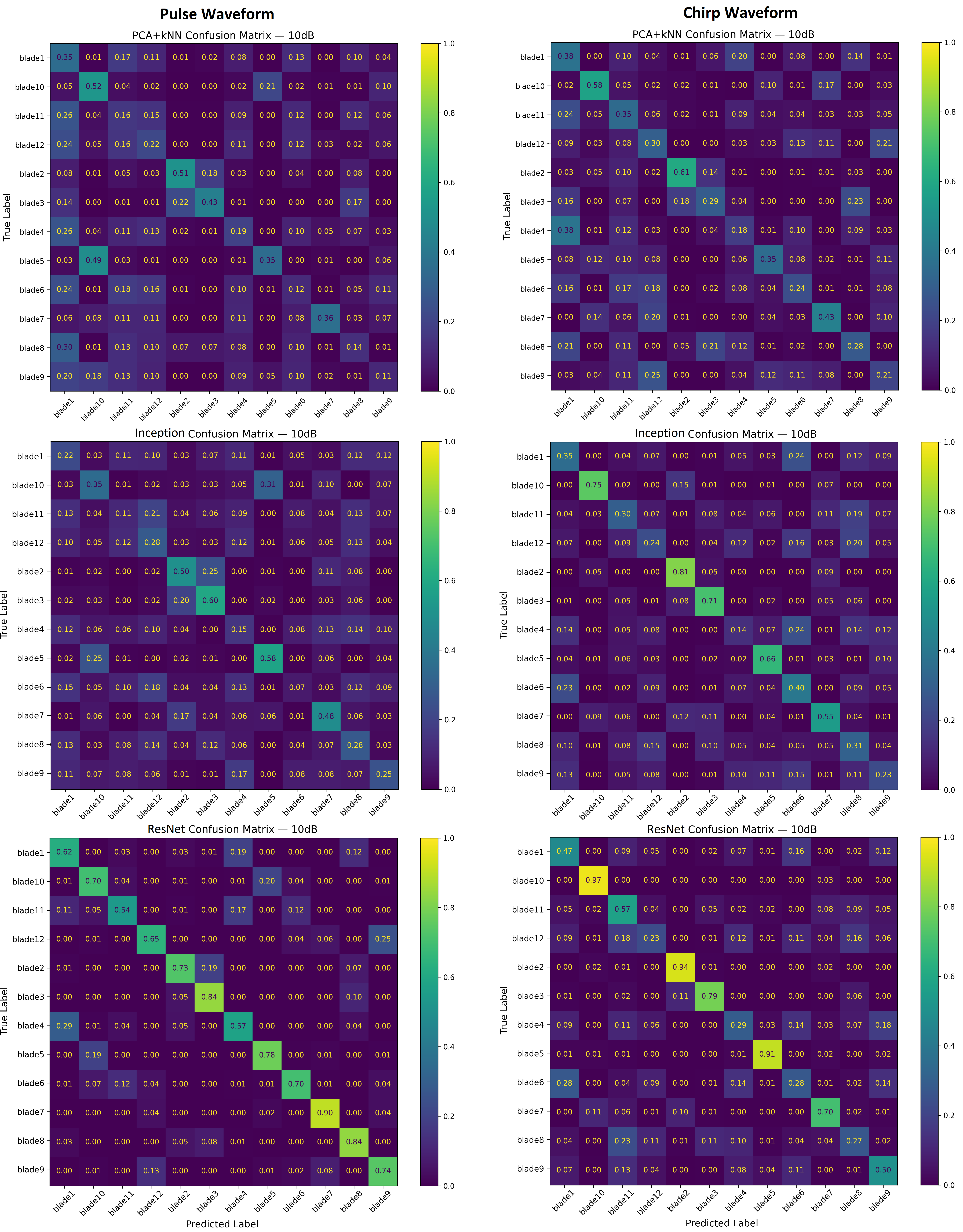}
\caption{Confusion matrices at 10 dB SNR for both the pulse (left-hand side) and chirp waveforms (right-hand side) from ADC0 on the first recording session (day1, chamber, ambient temperature). The first row shows a simple PCA + kNN classifier, followed by Inception and ResNet in the next two rows respectively}
\label{fig8}
\end{figure}

\end{document}